\documentclass{aa}
\usepackage{psfig}
\usepackage{lscape}
\usepackage{amsmath}
\usepackage{amssymb}
\usepackage[figuresleft]{rotating}

\newcommand{\MC}{\multicolumn}
\newcommand{\kms}{km~s$^\mathrm{-1}$}
\newcommand{\sunn}{$_{\odot}$}
\newcounter{qub}
\begin{document}

\title{
HS 2134+0400 - new very metal-poor galaxy, a representative of void
population?
}

\author{S.A. Pustilnik\inst{1}  \and
D. Engels\inst{2}      \and
A.Y. Kniazev\inst{3,1} \and
A.G. Pramskij\inst{1}  \and
A.V. Ugryumov\inst{1}  \and
H.-J. Hagen\inst{2}
}

\offprints{S. Pustilnik, \email{sap@sao.ru}}

\institute{
Special Astrophysical Observatory RAS, Nizhnij Arkhyz,
Karachai-Circassia,  369167 Russia
\and Hamburg Observatory, Gojenbergsweg 112, D-21029 Hamburg, Germany
\and European Southern Observatory, Karl-Schwarzschild Strasse 2,
   D-85748 Garching bei Muenchen, Germany
}

 \date{Received  July 18, 2005; accepted \hskip 2cm 2005}

\abstract{
We present the SAO 6\,m telescope spectroscopy of a blue compact galaxy (BCG)
HS~2134+0400 discovered in frame of the dedicated Hamburg/SAO survey
for Low Metallicity BCGs (HSS-LM). Its very low abundance of oxygen
(12+$\log$(O/H) = 7.44), as well as  other heavy elements (S, N, Ne, Ar),
assigns this dwarf galaxy to the group of BCGs with the lowest metal content.
There are only eight that low metallicity among several thousand known BCGs
in the nearby Universe. The abundance ratios for the heavy elements (S/O,
Ne/O, N/O, and Ar/O) are well consistent with the typical values of other
very metal-poor BCGs.
The global environment of HS~2134+0400 is atypical of the majority of BCGs.
The object falls within the Pegasus void, the large volume with the very low
density of galaxies with the normal (M$_{B}^{*}$=--19.6) or high luminosity.
Since we found in voids a dozen more the very metal-poor galaxies,
we discuss the hypothesis that such objects can be representative of a
substantial fraction of the void dwarf galaxy population.
   }

\authorrunning{S.A. Pustilnik et al.}

\titlerunning{HS 2134+0400: very metal-poor galaxy in a void}

   \maketitle

\section{Introduction}
\label{intro}

The heavy element contents in a galaxy interstellar medium (ISM)
is the parameter characterizing the current evolutionary state of a galaxy.
There is a general correlation between $Z$ - the metallicity of ISM (often
described in terms of the oxygen abundance O/H) and the galaxy mass, in
sense, that the massive, luminous galaxies (with luminosities of L$_{*}$ and
higher) have the characteristic values of O/H of about solar
value\footnote{According to the recent updates, the solar value of O/H is
accepted as 12+$\log$(O/H)=8.66 (Asplund et al. \cite{Solar04}).}
or a few times higher.
Low-mass, subluminous galaxies have rather wide O/H distribution, with the
great
majority of known O/H-values to be in the range of 1/10 to 1/2 of the solar
value (e.g., Kunth \& \"Ostlin \cite{Kunth2000} and references therein and
new data in Kniazev et al. \cite{SHOC}). In very small fraction of dwarf
galaxies ($\lesssim$2\%), the $Z$(ISM) is in the very low metallicity regime,
conditionally accepted as $Z < Z$\sunn/10, or 12+$\log$(O/H)
$\le$7.65 (e.g., Kunth \& \"Ostlin \cite{Kunth2000}). We call them
eXtremely Metal-Deficient (XMD) galaxies. The blue compact galaxy (BCG)
I~Zw~18 (Searle \& Sargent \cite{SS72}), with 12+$\log$(O/H)=7.17 (e.g.,
Izotov \& Thuan \cite{IT99}) was known during more than 30 years as the most
metal-poor object in the nearby Universe. Recently Izotov and Thuan
(\cite{IT05W}) have shown that another well known galaxy SBS 0335--052~W,
discovered in 1997 (Pustilnik et al. \cite{P97W}), is even more metal-poor,
with the oxygen abundance of 12+$\log$(O/H)=7.12.

Despite such galaxies are very rare in today's Universe, they are
interesting, on one hand, as the representatives of groups with less
typical evolution paths.
On the other hand, many processes in the ISM, including the star formation
(SF) and the interaction with massive stars, do strongly depend on the ISM
metallicity.
Therefore, the detailed study of the rare XMD galaxies can help in the
understanding of galaxy formation and their early evolution in the
high-redshift Universe, at the epochs when such galaxies were much more
typical.

The very low ISM metallicity in the local galaxies can be a result of various
evolutionary scenarios. They include: a) a significant metal loss due to a
powerful galactic superwind related to starbursts, b) the dilution of the
accumulated in the ISM metals due to the inflow of the intergalactic gas (or
the infall of intergalactic clouds) with the `pregalactic' metal content; c)
a very slow gas consumption rate and, respectively, a low metal production
rate in some galaxies, such as very low surface-brightness (LSB) galaxies;
and d) a very large delay in the SF onset, corresponding to the recent first
SF episodes in very stable protogalaxies.
Only the latter option results in a truly young galaxy in the local Universe.

Such interesting objects seem to exist among the known XMD galaxies, but only
a few reliable candidates for young galaxies are identified to date. The best
known candidate is the BCG I~Zw~18, for which Izotov \& Thuan (\cite{IT04})
have shown from the analysis of its very deep Hubble Space Telescope
colour-magnitude diagram  that the oldest stars in this galaxy
have ages $T \le$ 500 Myr (see also \"Ostlin \& Mouhcine \cite{Ostlin05}).
The other most probable candidate for young galaxies include first of all the
objects with the least known ISM metallicities (12+$\log$(O/H) $\le$ 7.29):
the components of the unique binary system (with the mutual
projected distance of 22 kpc) SBS 0335--052 E and W (Izotov et al.
\cite{Izotov99}, Lipovetsky et al \cite{Lipovetsky99}, Pustilnik et al.
\cite{VLA}, \cite{BTA}; Izotov \& Thuan \cite{IT05W}) and the nearest such
galaxy,  DDO~68 (Pustilnik et al. \cite{DDO68}).

To understand better the properties and the evolution status of XMD galaxies,
one needs in a sufficiently large sample of such objects. Despite their
paucity, there was a significant progress in the identification of new XMD
galaxies during the recent decade, mainly due to the large surveys of
emission-line galaxies, such as the Second Byurakan (SBS), Hamburg-SAO for
ELGs (HSS), Kitt Peak International Spectral (KISS), Hamburg-SAO for Low
Metallicity BCGs (HSS-LM). The significant number of new XMD galaxies is
also found among the objects of Sloan Digital Sky Survey (Kniazev et al.
\cite{Kniazev03}). Their current number is about a half hundred.

One of the surveys aimed at the search for new XMD galaxies is
the HSS-LM (Ugryumov et al. \cite{HSS-LM} (part I); Pustilnik et al. 2005,
in preparation (part II)). The four XMD galaxies from the paper of Ugryumov
et al. (\cite{HSS-LM}) are already studied and used, e.g., for the problem
of primordial helium (Izotov \& Thuan \cite{IT04}).
Here we report on a new galaxy HS~2134+0400 (J213658.95+041404.1), found in
the HSS-LM (part II), which belongs to the group of eight the most metal-poor
BCGs.
We present for this object the results of spectrophotometry, and the derived
physical parameters and element abundances. Discussing the properties of
HS~2134+0400, we emphasize its atypical spatial position, distant from the
cataloged luminous galaxies and their aggregates (the void environment), and
suggest the possible implications of finding in voids this and other
XMD galaxies.

\section{Observations and reduction}
\label{Obs}

The long-slit spectral observations were conducted with the SCORPIO
multi-mode instrument (Afanasiev \& Moiseev \cite{SCORPIO}),
installed in the prime focus of the SAO 6\,m telescope (BTA), during the
night of November 8, 2004.
The grism VPH550g was used with the 2K$\times$2K CCD detector EEV~42-40
and the exposed region of 2048$\times$600 px. This gave the range
$\sim$3500--7500~\AA\ with $\sim$2.0 \AA~pixel$^{-1}$ and FWHM $\sim$12~\AA\
along the dispersion.
The scale and the total extent along the slit were 0\farcs18 pixel$^{-1}$
and $\sim$2\arcmin, respectively. The slit with the position angle of
90\degr\  and the width of 1\arcsec\ (approximately along the major axis of
the main body, see Fig. \ref{fig:B_image}) crossed the bright central knot of
the galaxy. One 15-min exposure was obtained.
Before and after, the object spectrum  was complemented by the reference
spectra of He--Ne--Ar lamp for the wavelength calibration.
Bias and flat-field images were also acquired to perform the standard
reduction of 2D spectra. Spectral standard star Feige~34
(Bohlin \cite{Bohlin96}) was observed during the night for the flux
calibration.

The standard pipeline with the use of IRAF\footnote{IRAF: the Image Reduction
and Analysis Facility is
distributed by the National Optical Astronomy Observatory, which is
operated by the Association of Universities for Research in Astronomy,
In. (AURA) under cooperative agreement with the National Science
Foundation (NSF).}
and MIDAS\footnote{MIDAS is an acronym for the European Southern
Observatory package -- Munich Image Data Analysis System. }
was applied for the reduction of the long-slit spectra which included the
next steps.

Cosmic ray hits were removed from the 2D spectral frame in MIDAS.
Using IRAF packages from CCDRED, we subtracted bias and performed the
flat-field correction.
After that the 2D spectrum was wavelength-calibrated 
and the night sky background was subtracted.
Then, using the data on the spectrophotometry standard star,
the 2D spectrum was transformed to absolute fluxes.
One-dimensional spectrum of the central \ion{H}{ii} region was extracted by
summing up, without weighting, 11 rows along the slit ($\sim$2\arcsec), where
the principal line [\ion{O}{iii}]$\lambda$4363 for determination of
T$_{\rm e}$ was above the level of 2$\sigma_{\rm noise}$.

All emission lines were measured applying the MIDAS programs
described in detail in Kniazev et al. (\cite{SHOC}).
Briefly, they draw continuum, perform robust noise estimation and fit
separate lines by a single Gaussian superimposed on the continuum-subtracted
spectrum. Emission lines blended in pairs or triplets were fitted
simultaneously as blend of two or three Gaussians features. The quoted errors
of singular line intensities include the following components. The first is
related to Poisson statistics of a line photon flux. The second
component is the error resulting from the creation of the underlying
continuum, which gives the main contribution for the errors of faint lines.
For the fluxes of lines in blends, an additional error appears related to
the goodness of fit. The last term is related to the uncertainty of the
spectral sensitivity curve and gives an additional error for the relative
line intensities. For the presented observations this is  5\%, and, hence,
it gives the  main contribution
to the errors of the relative intensities of the strong lines.  All these
components were summed up squared. The total errors have been propagated to
calculate the errors of all derived parameters.

\begin{figure*}
   \vspace*{1cm}
   \centering
 \includegraphics[angle=-90,width=14cm, clip=]{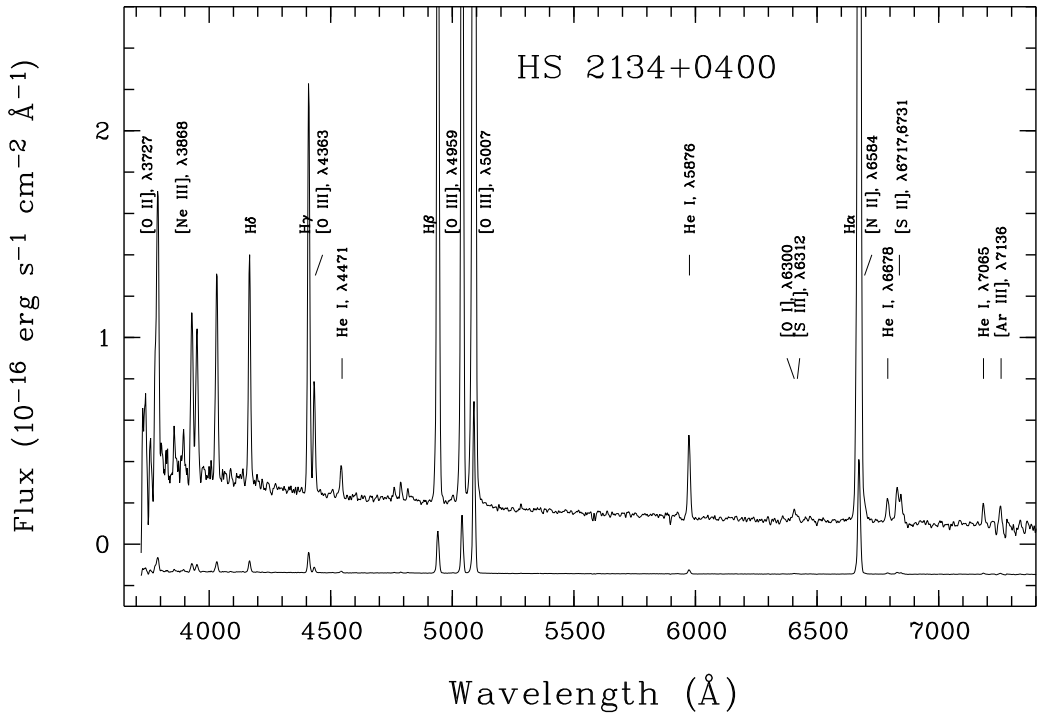}
   \caption{
    The spectrum of HS 2134+0400 with the  main lines marked.
    The same spectrum at the bottom is scaled down by a factor of 7
    to show the relative intensities of the strong lines.
       }
	 \label{fig:spectra}
 \end{figure*}

\begin{table}[hbtp]
\centering{
\caption{Line intensities and the derived parameters in the giant \ion{H}{ii}
region of HS 2134+0400}
\label{t:Intens}
\begin{tabular}{lcc} \hline  \hline
\rule{0pt}{10pt}
\rule{0pt}{10pt}
$\lambda_{0}$(\AA) Ion                    &
$F$($\lambda$)/$F$(H$\beta$)&$I$($\lambda$)/$I$(H$\beta$) \\ \hline

3727\ [O\ {\sc ii}]\            & 0.480$\pm$0.051 & 0.523$\pm$0.056   \\
3869\ [Ne\ {\sc iii}]\          & 0.221$\pm$0.018 & 0.238$\pm$0.020   \\
3967\ [Ne\ {\sc iii}]\ +\ H7\   & 0.251$\pm$0.016 & 0.268$\pm$0.025    \\
4101\ H$\delta$\                & 0.256$\pm$0.015 & 0.270$\pm$0.022   \\
4340\ H$\gamma$\                & 0.464$\pm$0.025 & 0.482$\pm$0.029   \\
4363\ [O\ {\sc iii}]\           & 0.126$\pm$0.010 & 0.131$\pm$0.010   \\
4471\ He\ {\sc i}\              & 0.038$\pm$0.006 & 0.039$\pm$0.006   \\
4686\ He\ {\sc ii}\             & 0.009$\pm$0.003 & 0.009$\pm$0.003   \\
4713\ [Ar\ {\sc iv]}\ +\ He\ {\sc i}\ & 0.019$\pm$0.005 & 0.020$\pm$0.005 \\
4740\ [Ar\ {\sc iv]}\           & 0.016$\pm$0.005 & 0.016$\pm$0.005   \\
4861\ H$\beta$\                 & 1.000$\pm$0.010 & 1.000$\pm$0.011   \\
4922\ He\ {\sc i}\              & 0.012$\pm$0.005 & 0.012$\pm$0.005   \\
4959\ [O\ {\sc iii}]\           & 1.307$\pm$0.068 & 1.298$\pm$0.068   \\
5007\ [O\ {\sc iii}]\           & 4.002$\pm$0.293 & 3.965$\pm$0.206   \\
5876\ He\ {\sc i}\              & 0.102$\pm$0.008 & 0.096$\pm$0.008   \\
6300\ [O\ {\sc i}]\             & 0.018$\pm$0.008 & 0.016$\pm$0.008   \\
6312\ [S\ {\sc iii}]\           & 0.007$\pm$0.004 & 0.006$\pm$0.004   \\
6548\ [N\ {\sc ii}]\            & 0.006$\pm$0.006 & 0.006$\pm$0.006   \\
6563\ H$\alpha$\                & 3.004$\pm$0.153 & 2.740$\pm$0.152   \\
6584\ [N\ {\sc ii}]\            & 0.020$\pm$0.006 & 0.019$\pm$0.006   \\
6678\ He\ {\sc i}\              & 0.030$\pm$0.007 & 0.027$\pm$0.006   \\
6717\ [S\ {\sc ii}]\            & 0.048$\pm$0.009 & 0.044$\pm$0.008   \\
6731\ [S\ {\sc ii}]\            & 0.040$\pm$0.009 & 0.036$\pm$0.008   \\
7065\ He\ {\sc i}\              & 0.022$\pm$0.005 & 0.020$\pm$0.005   \\
7136\ [Ar\ {\sc iii}]\          & 0.022$\pm$0.006 & 0.019$\pm$0.005    \\
C(H$\beta$)\ dex          & \MC {2}{c}{0.12$\pm$0.07}   \\
EW(abs)\ \AA\             & \MC {2}{c}{0.00$\pm$2.27}   \\
$F$(H$\beta$)$^a$\        & \MC {2}{c}{45.8$\pm$0.6}    \\
EW(H$\beta$)\ \AA\        & \MC {2}{c}{ 214$\pm$11}    \\
Rad. vel.\ \kms\          & \MC {2}{c}{5070$\pm$45}     \\
\hline  \hline
\MC{3}{l}{$^a$ in units of 10$^{-16}$ ergs\ s$^{-1}$cm$^{-2}$.}\\
\end{tabular}
 }
\end{table}

\section{Results}
\label{results}

\subsection{Line intensities and element abundances}
\label{abun}

Relative intensities of all emission lines used for the abundance
determination in the giant \ion{H}{ii} region of HS~2134+0400,
as well as the derived C(H$\beta$), EWs of Balmer absorption lines, the
measured flux in H$\beta$ emission line and the measured heliocentric
radial velocity are given in Table \ref{t:Intens}.
The 1D spectrum of the object is shown in Figure~\ref{fig:spectra}.
Extinction is low: C(H$\beta$) $\sim$0.1, what is well consistent with the
extinction data for the majority of very metal-poor galaxies.

\begin{table}[hbtp]
\centering{
\caption{Element abundances in HS~2134+0400}
\label{t:Chem}
\begin{tabular}{lc} \hline  \hline
$T_{\rm e}$(OIII)(10$^{3}$~K)\             & 19.81$\pm$1.04      \\
$T_{\rm e}$(OII)(10$^{3}$~K)\              & 15.63$\pm$0.76      \\
$T_{\rm e}$(SIII)(10$^{3}$~K)\             & 18.14$\pm$0.86      \\
$N_{\rm e}$(SII)(cm$^{-3}$)\               & 236$\pm$570~        \\
&  \\
O$^{+}$/H$^{+}$($\times$10$^{-5}$)\        & 0.411$\pm$0.067     \\
O$^{++}$/H$^{+}$($\times$10$^{-5}$)\       & 2.310$\pm$0.290     \\
O$^{+++}$/H$^{+}$($\times$10$^{-5}$)\      & 0.027$\pm$0.012     \\
O/H($\times$10$^{-5}$)\                    & 2.748$\pm$0.298     \\
12+$\log$(O/H)\                            & ~7.44$\pm$0.05~     \\
& \\
 N$^{+}$/H$^{+}$($\times$10$^{-6}$)\        & 0.129$\pm$0.033~~     \\
 ICF(N)\                                    &  6.686                 \\
 $\log$(N/O)\                               & --1.50$\pm$0.12~~     \\
 &  \\
Ne$^{++}$/H$^{+}$($\times$10$^{-5}$)\      & 0.283$\pm$0.039     \\
ICF(Ne)\                                   & 1.190                 \\
$\log$(Ne/O)\                              & --0.91$\pm$0.08~     \\
& \\
S$^{+}$/H$^{+}$($\times$10$^{-7}$)\        & 0.747$\pm$0.126~~     \\
S$^{++}$/H$^{+}$($\times$10$^{-7}$)\       & 1.930$\pm$1.107~~     \\
ICF(S)\                                    & 1.865                  \\
$\log$(S/O)\                               & --1.74$\pm$0.19~~     \\
& \\
Ar$^{++}$/H$^{+}$($\times$10$^{-7}$)\      & 0.519$\pm$0.143~~     \\
Ar$^{+++}$/H$^{+}$($\times$10$^{-7}$)\     & 1.267$\pm$0.435~~     \\
ICF(Ar)\                                   & 1.019                  \\
$\log$(Ar/O)\                              & --2.18$\pm$0.12~~     \\
\hline   \hline
\end{tabular}
 }
\end{table}

To derive the element  abundances of species O, Ne, N, S, and Ar in the
giant \ion{H}{ii} region of HS 2134+0400, we use
the standard method from Aller (\cite{Aller84}), and follow the procedure
described in detail by Pagel et al. (\cite{Pagel92}) and Izotov et al.
(\cite{ITL97}).
Chemical abundances and physical parameters are determined in the frame of
the classical  two-zone model of \ion{H}{ii} region (Stasi\'nska
\cite{Stas90}), as described in detail in our recent papers (Pustilnik
et al. \cite{HS0837},  Kniazev et al. (\cite{SHOC}; \cite{Sextans}).
The derived  electron temperatures T$_{\rm e}$  and density N$_{\rm e}$,
as well as the abundances of oxygen, nitrogen, neon, sulfur, and argon
are given in Table~\ref{t:Chem}. The resulting heavy element abundances (in
particular, the value of 12+$\log$(O/H)=7.44$\pm$0.05) assign this BCG as the
eighth most metal-poor galaxy of several thousand known BCG/\ion{H}{ii}
galaxies in the nearby Universe.

\section{Discussion and Conclusions}
\label{discussion}

\subsection{HS 2134+0400 in comparison to other very metal-poor BCGs}

The element abundance ratios of Ne/O, S/O, N/O and Ar/O for HS~2134+0400
are consistent within (1.0--1.5)$\sigma$ uncertainties with the abundance
patterns
for the group of the most metal-poor BCGs (Izotov \& Thuan \cite{IT99}).
This fact is consistent with the production of all these elements in the
XMD BCGs within the same massive stars along with oxygen.

\begin{figure}
   \vspace*{1cm}
   \centering
 \includegraphics[angle=0,width=8cm, clip=]{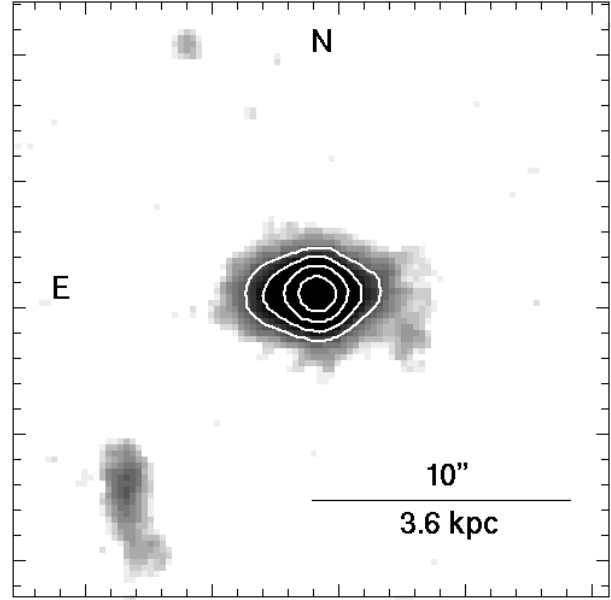}
   \caption{
    $B$-band image of HS 2134+0400, obtained with the Wide Field Imager
    at the ESO 2.2\,m telescope (see details in the forthcoming paper by
    Pustilnik et al. 2005b). Isophotes with the step of 1$^{m}/\Box{\arcsec}$
    show the
    structure of the innermost region. At the adopted distance of 74 Mpc,
    the BCG East-West extent of $\sim$9\arcsec\ corresponds to 3.2 kpc.
       }
	 \label{fig:B_image}
 \end{figure}

The optical morphology of this faint galaxy ($B_{\rm tot}$=19\fm3, Pustilnik
et al. 2005, in prep.) with the bright central SF region, dominating the
galaxy light, is quite typical of BCGs (Fig. \ref{fig:B_image}). The outer
isophotes are rather irregular and disturbed. There is a curved plume on the
western edge. According to Telles et al. (\cite{TT}) and Bergvall \&
\"Ostlin (\cite{Bergvall02}), such morphology is preferably related to the
luminous BCGs (M$_{\rm B}$ of\ --17 to\ --20) and is suggested to be caused
by strong interactions, and especially by mergers.
Despite this BCG is not a very luminous (M$_{B}^{0}$ = --15\fm1), its
merger nature is not excluded. To check this option, more detailed studies
are necessary.
In particular, several BCGs with M$_{B}^{0}$ as low as --14\fm5 to --15\fm4 in
the sample of 86 BCGs from the zone of the Second Byurakan Survey (Pustilnik
et al. \cite{PKLU}) were identified as having the `merger' morphology.
Moreover, some other XMD galaxies are clearly
related to the merger phenomenon. Some of them are also not luminous.
They include, in particular,  Dw~1225+0152 ($M_{\rm B} =$ --15; Salzer et al.
\cite{Salzer91}; Chengalur et al. \cite{Chengalur95}), HS 0837+4717
(Pustilnik et al. \cite{HS0837}), SBS~0335--052 E and W (Pustilnik et al.
\cite{VLA, BTA}).

According to the starburst models of Schaerer \& Vacca (\cite{SV98}) for the
respective metallicity of $z$=0.001 and the standard Salpeter IMF, the
measured value of EW(H$\beta$)=214~\AA\
implies the age of the instantaneous SF burst of 3.2~Myr that corresponds
to the maximum of Wolf-Rayet (WR) phase at this metallicity. This suggests
that the higher
S/N spectra can reveal the characteristic `blue' WR bump and determine the
fraction of WR stars relative to the full number of massive stars. This is
an important observational parameter to test the models of massive star
evolution in the very metal-poor environment.

Various evolutionary scenarios can lead to that small ISM metallicities. In
particular, very low SF rate over the whole galaxy life will result in the
small gas consumption and little ISM enrichment. Such
a situation is expected in very low surface brightness galaxies.
Indeed, almost all studied LSB dwarf galaxies with the oxygen abundances of
12+$\log$(O/H) $\le$7.50 (UGCA 20, UGCA 292, UGC 2684, AM~0624--261) appeared
to have the old red stellar population (van Zee et al. \cite{vZee_UA20},
van Zee \cite{vZee_colours}, Parodi et al. \cite{Parodi02}),
thus, they are old objects. On the other hand, several
BCGs (or dIrr/BCG) with the lowest metallicities show no tracers of the old
stellar population (I~Zw~18, SBS~0335--052~E and W) and are currently
considered as the best candidates for the genuine local young galaxies.
By analogy, HS~2134+0400, as one of the lowest metallicity BCGs, also can
belong to this group and therefore it deserves a more detailed
multiwavelength study.

\subsection{HS 2134+0400 as a `void' galaxy}
\label{void}

As shown, e.g., by Salzer (\cite{Salzer89}), Pustilnik et al.
(\cite{PULTG}), and Popescu et al. (\cite{Popescu97}),
BCGs and \ion{H}{ii} galaxies, as subluminous objects,
follow in general the spatial distribution of luminous galaxies
(L $\ge$ L$_{*}$). However,  they show more scattering around the structures
delineated by luminous galaxies, and do escape the densest regions.
Only a small fraction of BCGs (15$\pm$5~\%) fills in voids, the large regions
completely devoid of luminous galaxies. Other subluminous late-type galaxies
also partly fill in voids, forming some filamentary structures (e.g., Lindner
et al. \cite{Lindner96}).

This new XMD BCG is situated in the region of very low density of
luminous galaxies (M$_B$ $\le$M$_B^{*}$=--19.6, for the accepted here
H$_{0}$ = 72~\kms~Mpc$^{-1}$). This region belongs to the outskirts of the
huge Pegasus void, described by Fairall (\cite{Fairall98}). This void
is characterized by the coordinates of its center (B1950) RA=22$^{h}$,
Dec = +15\degr,  $cz$ = 5500~\kms,  and by the void diameter of 3000~\kms\
($\sim$40 Mpc).
HS 2134+0400 is quite far from the void center, closer to its SW rim, not
far from its foreground edge. We have examined the environment of this
galaxy using the Updated Zwicky Catalog redshifts (Falco et al.
\cite{Falco99}) and the additional information on fainter galaxies from NED.
The nearest object to HS~2134+0400  is the luminous galaxy pair CGCG 402-013
at D=4.0 Mpc (NGC 7102, $M_{\rm B}$ = --19.7).
The next nearest galaxies are CGCG 402-004 at 4.4 Mpc ($M_{\rm B}$
= --18.5)
and UGC~11792 at 4.8 Mpc ($M_{\rm B}$ = --18.6).
All other known galaxies are at the distances more than 6 Mpc.
It is worth noting that HS~2134+0400 is not the only XMD galaxy found in
the Pegasus void. The BCG HS~2236+1344 with 12+$\log$(O/H)=7.47 (Ugryumov
et al. \cite{HSS-LM}, Izotov \& Thuan \cite{IT04}) also falls in this void.
It is situated on the opposite side of the void, somewhat closer to its
center.

How common are very low metallicity galaxies in voids? The quantitative
comparison of the metal content of dwarf galaxies in voids and in the denser
environment still waits for the good quality O/H determinations
performed for large well selected samples. Below we mention several other
examples of a dozen XMD galaxies found by us in voids (paper in
preparation), which probably will stimulate the studies of this issue.
In particular, in one of the nearest small voids, the Lynx-Cancer one, having
a diameter of $\sim$800~\kms,  three XMD galaxies are discovered with
12+$\log$(O/H) from 7.21 to 7.65 (Pustilnik et al. \cite{SAO0822,DDO68}).
Several other known XMD BCGs, like HS 0837+4717 (Pustilnik
et al. \cite{HS0837}) also fall in voids delineated by luminous galaxies.

One more interesting example is that of two XMD galaxies falling close to
the center of a known void. These are HS~1313+4521 with 12+$\log$(O/H)=7.57
and V$_{\rm hel}$=3449~\kms\ (Pustilnik et al. 2005c, in preparation)
and KISSR~1490 (1311+4418) with 12+$\log$(O/H)=7.56 and V$_{\rm hel}$ =
3559~\kms\ (Lee et al. \cite{Lee04}). The both dwarf galaxies
are situated within $\sim$3 Mpc from the center of the void `C4' defined by
Lindner et al. (\cite{Lindner95}). Its center coordinates (B1950) correspond
to RA = 13$^{h}$19$^{m}$, Dec = +47.5\degr, $cz$ = 3690~\kms, its diameter
corresponds to 1500~\kms\ ($\sim$20 Mpc).

If the appearance of many XMD galaxies in voids is not occasional, whether
some natural explanations for such a correlation can be suggested? Probably
yes. Despite of such cases were not described in the literature at that time,
Peebles (\cite{Peebles01}) suggested that unusual objects similar to
SBS~0335--052 would be natural to find in voids. Our discoveries
of many XMD galaxies in voids seems proves his prediction.
In the frame of the Cold Dark Matter cosmology, the galaxies formed in
voids, should be predominantly
of lower mass and retarded in their formation epoch in comparison to the
galaxies from the average and higher density regions (e.g., Gottl\"ober et
al. \cite{Gott03}).
In the frame of the popular hierarchical galaxy formation scenario
(e.g., White \& Frenk \cite{WF91}), due to the reduced galaxy density
in voids, a part of these dwarfs can belong to the group of galaxies,
experienced very few, or even no collisions during their life. They will
differ quite much from the more common galaxies which past through many
interactions and/or mergers.
In particular, one could expect that a fraction of void galaxies can
survive in their nascent state, avoiding any significant interaction that
triggers intensive star-forming episodes and causes the additional ISM metal
enrichment.

\subsection{Conclusions}

\label{Conclusions}

Summarizing the presented observational results and the discussion above, we
draw the following conclusions:
\begin{itemize}
\item The oxygen abundance in the giant \ion{H}{ii} region of the blue compact
      galaxy HS~2134+0400 is of 12+$\log$(O/H)=7.44$\pm$0.05 dex,
      it belongs to the eight  most metal-poor BCGs.
\item  The heavy element abundance ratios  S/O, Ne/O, N/O and Ar/O are well
       consistent with the average values derived on the small group of the
       most metal-poor  BCGs.
\item  HS~2134+0400 falls to the region with the low density of
       luminous  galaxies, at the outskirts of the large Pegasus void, in
       which one more very metal-poor BCG HS 2236+1344 is found.
\item  These two objects, along with many other XMD galaxies discovered in
       voids, may comprise a representative group of the void galaxy
       population. The existence of such a group would indicate
       the importance of the effect of global environment on the chemical
       evolution of low-mass galaxies.
\end{itemize}

\begin{acknowledgements}

The authors thank A.V.~Moiseev for the help in observations with SCORPIO.
SAP, AGP and AVU acknowledge the partial support from Russian state program
"Astronomy". This research has made use of the NASA/IPAC Extragalactic
Database (NED) which is operated by the Jet Propulsion Laboratory,
California Institute of Technology, under contract with the National
Aeronautics and Space Administration.

\end{acknowledgements}

\renewcommand{\baselinestretch}{0.5}

\end{document}